\documentclass[12pt]{article}

\def\s{\sigma}

\def\k{\kappa}

\def\d{\delta}
\def\r{\rho}
\def\h{\hat}

\def\dda{\ddot{a}}

\def\l{\label}

\def\m{\mu}

\def\be{\begin{equation}}
\def\ee{\end{equation}}
\def\ba{\begin{eqnarray}}
\def\ea{\end{eqnarray}}  

\def\h{\hat}

\def\t{\tilde}
\begin{document}

\vspace{4cm}
\centerline{\large\bf A note on cosmology in a brane model}
\vspace{2cm}
\centerline{\bf Mikhail Z. Iofa \footnote{ e-mail:
iofa@theory.sinp.msu.ru}} \centerline{Skobeltsyn Institute of Nuclear
Physics}
\centerline{Moscow State University}
\centerline{Moscow 119992, Russia}  

\begin{abstract}
We study some aspects of cosmology in a five-dimensional
model with matter, radiation
and cosmological constant
on the four-dimensional brane(s) and without matter in the bulk. 
The action of the model does not contain explicit curvature terms on the 
the brane(s).
We obtain 
solution of the generalized Friedman equation as a function of dimensionless
ratio of the scales 
$b^2 =
\frac{\mu M^2_{pl}}{M^3}$ ( $\mu$ is the scale in the warp factor in the 5D
metric which is
taken of order $10^{3\div 4}GeV$, $M\geq\mu$ is the $5D$ fundamental scale
). We  assume that there is a hierarchy between  4D and
5D scales.  For $b^2 =O(1)$ the age of the Universe is
found comparable, but below the current experimental value, for $b^2\gg 1$
it is obtained  much smaller than  the experimental bound.  
Because time dependence of temperature of the Universe
in the 5D model is different from that in the standard cosmology, 
the abundance of ${}^4$He produced
in the primordial nucleosynthesis is 
obtained about three times more than in the standard cosmology. 
\end{abstract}   
\newpage
Recently there has been much activity  in the study of the higher-dimensional 
models as in
connection with possible detection of  hypothetical large extra dimensions
at the future colliders, so as with a
possibility of solution of the hierarchy problem \cite{rub,csaki,rizzo,sund}.  
If considered as models
of the physical
world,  these models pose a
question of possible cosmologies. 
In this note we consider a five-dimensional model with the visible world
located on a 3-brane embedded in the $AdS_5$ space-time.

We study some aspects of cosmology in a model with matter and
cosmological constant(s) on
the brane(s) and with cosmological constant, but without matter in the bulk.
 In 5D models the form of the Friedman
equation on 3-branes is
different from that in the standard cosmology \cite{BDL1,BDL2,SMS} and
contains
 terms quadratic in components of the energy-momentum tensor on a brane.

The energy-momentum tensor on the visible brane is taken in the form
\be  
\l{2}
T_A^B=\d (y)\frac{\sqrt{-g^{(4)}}}{\sqrt{-g^{(5)}}}
\left[- diag(\h\s ,\h\s ,\h\s ,\h\s ,0 )+
diag(-\h\r , \h p ,\h p , \h p, 0 )\right]
,\ee
where $\h\r (t) $ and $\h p (t) $ are the sum of densities and pressures of
cold matter and radiation and $\h{\s}$ is the cosmological constant on the
brane. The visible brane is located at the fixed position at $y=0$.

We consider a class of solutions to the Einstein equations in 5D model with the metrics
of the form
\be
\l{1}
ds_5^2 =g_{AB}dx^A dx^B =-n^2 (t,y)dt^2  +a^2 (t,y)\gamma_{ij}dx^i dx^j
+dy^2 .
\ee
Solution of the
Einstein equations is assumed to be invariant under the symmetry
$y\rightarrow -y$. 

The  fundamental 4D and 5D scales are $M_{pl}$ and $M$, the
gravitational
constants are $\k^2_4 ={8\pi}/{M^2_{pl}}$ and $\k^2 ={8\pi}/{M^3}$.
The characteristic scale of the warp factor of the metric in the bulk is
$\mu =\left(\k^2\Lambda/6\right )^{1/2}$,
 where $\Lambda$ is the cosmological constant in the bulk.

Possible cosmologies are strongly dependent on whether the curvature
terms on 3-brane(s) are introduced explicitly  in the 5D action or not.
Cosmologies in the models with curvature terms on the branes were
investigated in papers \cite{dva,Shtanov1,Shtanov2,Shtanov3,Def}.
In this note we consider a model without 
 explicit 4D curvature terms in the 5D action.

We study cosmologies by solving the generalized Friedman equation. 
The age of the Universe is calculated as a function of 
dimensionless ratio of the scales $b^2 \equiv \mu M^2_{pl} /M^3$.
We assume that there is a hierarchy between the Planck and 5D
scales $M_{pl}$ and $M$.

Solving Friedman equation we find time dependence of the scale factor 
and of temperature of the
Universe. In the 5D model this dependence is different from that
in the standard 4D cosmology. This leads, in particular, to different rates of
primordial nucleosynthesis in both cases. We calculate the abundance of 
${}^4$He and compare it with that in the standard cosmology. 

Let us introduce the normalized expressions for energy density, pressure and
cosmological constant on the visible brane which all have the same
dimensionality $GeV$
\be  
\l{4}
\quad \r =\frac{\k^2\h\r}{6}, 
\quad \s =\frac{\k^2\h\s}{6},\quad {p} =\frac{\k^2\h{p}}{6}.
\ee
From the Einstein equations and the junction conditions for the
functions $a(y,t)$ and $n(y,t)$ on the 
branes is derived the generalized 
Friedman equation. 
Using the freedom in the
choice of parametrization of space-time variables, the function $n(y,t)$ on
the visible brane is made constant on the visible brane
$$
n(0,t) =1.
$$
We study cosmology by solving the generalized Friedman
equation on the visible brane in the form
with the second-order time derivatives of the scale factor \cite{BDL1,SMS}
\be
\l{3}
\frac{\ddot{a}(0,t)}{a(0,t)} +\frac{\dot{a}^2 (0,t)}{a^2 (0,t)} = -2\mu^2 -  
\left(\s +\sum \r_i \right)\left(\s +\sum \r_i +3 (-\s +\sum p_i
)\right)
,\ee
where  $(\r_i ,\,\,p_i ) =(\r_m ,\, p_m) \,\, (\r_r ,\, p_r )$
are  densities and pressures of
the total cold matter and radiation.

In terms of  cosmological variable
$$ 
z =\frac{a(0, t_0 )}{a(0, t)}-1
,$$  
where  $a(0 ,t_0 )$ is the current-time scale factor, $z$-dependencies of
 cold matter
and radiation  densities have the following form
\be
\l{5}
\h{\r}_m =\h{\r}_{m0} (1+z)^3, \quad \h{\r}_r =\h{\r}_{r0} (1+z)^4
.\ee
As the input we use the current-time values of the  Hubble parameter
$H_0\sim
10^{-42} GeV$ and  cold matter and radiation densities, or
 $\Omega_m =\h{\r}_{m0}/\h{\r}_C \sim 0.2 \div 0.4$ and $\Omega_r =
\h{\r}_{r0}/\h{\r}_C \sim 10^{-4} $, 
where $\h{\r}_C$ is the
critical density $\h{\r}_C \equiv 3H^2_0 M^2_{pl} /8\pi $ \cite{kolb,wein,part}.
The normalized matter density can be written as
$$
\r_{m0} =\frac{b^2 H^2_0\Omega_m}{2\mu}
.$$
We assume that $\mu \sim 10^{3\div 4}GeV$ and $M\geq \mu$ (cf. \cite{rub}).
First, let us consider the case $1\leq b^2 <10^{32}$,
the upper limit corresponding to $M\sim
\mu\sim 10^{3\div 4}GeV$.
Rearranging the terms in the rhs of
(\ref{3}), we rewrite it as
\footnote{Having in view a model of the RSI type \cite{RS1,cs4,cs41} with matter, below we take
a negative cosmological constant on the visible brane. The "effective"
cosmological constant $\m -|\s|$ is small and positive.}
\be
\l{6}
\frac{\dda(0,t)}{a(0,t)} +\frac{\dot{a}^2 (0,t)}{a^2 (0,t)} =2(\s^2 -\mu^2
)
-|\s|\sum (\r_i -3p_i )-\sum \r_i \sum (\r_j +3p_j ).
\ee
Introducing new variable
$$
u(t)=(z+1)^{-2}=\left({a(0, t)}/{a(0, t_0 )}\right)^2,
$$
 we obtain the first
integral of Eq.(\ref{6}) in the form
\be
\l{7}
(\dot{u})^2 =4\left[(\s^2 -\mu^2 )u^2 -2|\s_1| \r_{m0}u^{1/2}
+\r_{m0}^2u^{-1} +2\r_{m0}\r_{r0}u^{-3/2}+\r_{r0}^2 u^{-2} +C\right]
,\ee
where $C$ is an integration constant. 

Setting in (\ref{7}) $t=t_0$ and substituting  $\dot{u}/{u}|_{t=t_0}=2H_0$
 we obtain
\be
\l{8}
H^2_0 =\s^2 -\mu^2 - 2|\s_1| \r_{m0} +(\r_{m0} +\r_{r0} )^2 +C
.\ee
The second relation is obtained by taking (\ref{6}) at $t=t_0$
  and writing $\dda(t_0 )/a(t_0 )=-q_0 H^2_0$,
where $|q_0|=O(1)$ is the deceleration parameter,
\be
\l{8a}
(1-q_0 )H^2_0 =2(\s^2 -\mu^2 ) - |\s| \r_{m0} -(\r_{m0}+\r_{r0})\r_{m0}
.\ee

Let us introduce the ratio 
\be
\l{81}
r^2 =\frac{H^2_0}{ |\s|\r_{m0}} = \frac{2}{b^2 \Omega_m
}\frac{\mu}{|\s|}.
\ee
From Eqs. (\ref{8}) and (\ref{8a}), neglecting the term quadratic in
densities,  we obtain
\be
\l{8b}
 C= |\s|\r_{m0}\left(\frac{3}{2} +\frac{1+q_0 }{2}r^2\right) ,\qquad
\s^2 -\mu^2= |\s|\r_{m0}\left(\frac{1}{2} +\frac{1-q_0 }{2}r^2\right)
.\ee
Solving the second equation (\ref{8b}) with respect to $|\s|$, we
have
$$
|\s|\simeq \mu +\frac{\r_{m0}}{4} +\frac{1-q_0}{4\mu}H^2_0.
$$
Thus, $|\s|\simeq \mu$ and $r^2\simeq 2/b^2 \Omega_m $.

Substituting in Eq.(\ref{7}) the equality $|\s|\r_{m0}={H^2_0}/{r^2}$
which follows from the definition (\ref{81}) and
setting  $|\s|\simeq \mu$, we  rewrite  Eq.(\ref{7}) in a form
\be
\l{9}
(\dot{u})^2 =4\frac{H^2_0}{r^2}\left[\frac{1+(1-q_0 )r^2}{2}u^2-2u^{1/2}+
\frac{H^2_0}{r^2 \m^2}u^{-1}\left(1+\frac{\r_{r0}}{\r_{m0}}
u^{-1/2}\right)^2 +
\frac{3+(1+q_0 )r^2}{2}\right].   
\ee
For $u\ll 1$ the term quadratic in densities , estimated with $\m\sim
10^{3\div 4} GeV$, $ {\r_{r0}}/{\r_{m0}}\simeq 2\cdot 10^{-4}$
\cite{part}   
 and $H_0\sim
10^{-42}GeV$  , is of order the last constant term in (\ref{9})
for $u< 10^{-(49\div 50)}r^{-1}$, or $z\geq 10^{25}b^{1/2}$.    

To obtain a qualitative picture and to estimate the age of the Universe,
 we consider   the following
regions of $u$ (or $z+1 =u^{-1/2}$).
\newline
(i) Cold matter-dominated region
$1>u>\left({\r_{r0}}/{\r_{m0}}\right)^2\sim 10^{-8}$ or $1\leq z <10^4$,
where  Eq.(\ref{9}) can be approximated as
\be
\l{r1}
\dot{u}^2 \simeq 4\frac{H^2_0}{r^2}\left[\left
(\frac{1}{2} + \frac{(1-q_0 )r^2}{2} \right)u^2
-2u^{1/2} + \left(\frac{3}{2} +\frac{ (1+q_0 )r^2}{2}\right )\right],
\ee
(ii)Radiation-dominated region where all
$u$-dependent terms are small as compared to the constant term
\footnote{At the point 
of transition from the matter-dominated to the
radiation-dominated phase  $z\sim 10^4$  the terms quadratic in
densities are of the same order.}
$$
10^{-8}>u, \hspace*{0.5cm}\qquad \frac{3+(1+q_0 )r^2}{2}>u^{-2}
\frac{H^2_0}{r^2 \m^2}
\left(\frac{\r_{r0}}{\r_{m0}}\right)^2 ,
$$
or $10^{-8}>u>10^{-(49\div 50)} /r^2$.
In this region Eq.(\ref{9}) can be written as
\be                             
\l{r2}                          
\dot{u}^2 \simeq H^2_0 \frac{6 +2(1+q_0 )r^2 }{r^2}.
\ee         
(iii)       
Radiation-dominated high-energy  region $u<10^{-50}b$, or
$z>10^{25}r^{1/2}$,
where Eq.(\ref{9}) can be approximated as
\be                                        
\l{r3}                                     
\dot{u}^2 \simeq 4\r_{r0}^2 u^{-2}.         
\ee                                        
In both regions (ii) and (iii) we use the approximate equation
\be              
\l{r4}           
\dot{u}^2 \simeq 4\left[\frac{3+(1+q_0 )r^2 }{2}
 \m\r_{m0} +\r_{r0}^2 u^{-2}\right].
\ee     

To estimate the age of the Universe we can consider only
the matter-dominated period.  
Integrating  Eq.(\ref{r1}), we obtain
\be
\l{r7}
 H_0 (t_0 -t)
=\frac{r}{\sqrt{2}} \int^1_{u (t)}\frac{du}
{\left[(1+(1-q_0 )r^2 )u^2 -4u^{1/2} +3+(1+q_0 )r^2\right]^{1/2}}.
\ee
For an estimate we can take $t=0$ and $u(t)=0$.
 
First, let us suppose  that both scales $\m$ and $M$ are of
order $10^{3\div 4}GeV$. In this case $b^2\sim 10^{32}$ and
 $r^2\sim 10^{-(30\div 32)}$. 
For small $r^2$  the
integral (\ref{r7}) can be written as
\be
\l{r5}
 H_0 t_0 
\simeq r\sqrt{2}\int_{0}^{1}\frac{vdv}
{\left[v^4 -4v +3 +r^2 ((1-q_0)v^4 +1+q_0 )\right]^{1/2}}
\simeq 
r\int_{0}^{1}\frac{vdv}{\left[3(v-1)^2 +r^2 \right]^{1/2}}
\ee
Here we have set $v^4 -4v +3 =(v-1)^2 (v^2 +2v+3)\simeq 6(v-1)^2$ and
$(1-q_0)v^4 +1+q_0 \simeq 2 $.
For $r^2 \ll 1$ the age of the Universe
\be     
\l{r6}  
t_0 \simeq \frac{r}{\sqrt{3}H_0}\left[\ln\frac{2\sqrt{3}}{r} -1 \right]
\ee
 is much  smaller than that in the standard cosmology.

Next, we take $\m\sim 10^{3\div 4}GeV$,  keeping the 5D scale $M$ or,
equivalently, 
$r^2$  as  free parameters.
The age of the Universe $t_0$ calculated with $q_0=0,\,\pm 1/2$
and $r^2 =O(1)$ is
\begin{center}
\begin{tabular}{|c|c|c|c|c|c|c|c|c|c|c|c|} \hline
$ r^2 =2/b^2\Omega_m = $&
0.001 & 0.01 & 0.05 & 0.2 & 0.4 & 0.8 & 1.0 & 1.4 & 2.0 & 4.0 & 10  \\
\hline
$q_0=+1/2,\,\, t_0 H_0 =$&
0.07  & 0.16  & 0.26 & 0.37 & 0.42 & 0.46 & 0.47 & 0.49 & 0.50 & 0.52 &
0.54 \\ \hline
$\hspace{0mm}q_0 =0,\qquad t_0 H_0 = $&
0.07  & 0.16 & 0.27 & 0.38 & 0.44 & 0.49 & 0.51 & 0.53 & 0.55 & 0.58 & 0.60
\\  \hline
$q_0 = -1/2,\,\,\,\, t_0 H_0 =$&
0.07 & 0.16 &0.27  &0.40 &0.47 & 0.54 & 0.56 & 0.59 & 0.62 & 0.67   & 0.72
\\ \hline
\end{tabular}
\end{center}

For $r^2\sim 1\div10$  the age of the Universe
is comparable, but smaller than that in the   
standard cosmology in the $\Lambda$CDM model
(~$t_0\sim 0.65 H^{-1}_0$  
in matter-dominated model with
$\Omega_m =1$ and $\Lambda =0$, and $ t_0\simeq  H^{-1}_0$ in the $\Lambda$CDM
model with $\Omega_m =0.28$ and $\Omega_m +\Omega_\Lambda =1 
$ \cite{kolb, part,sah}).  
With  
 $b^2 =O(1)$ and $\mu\sim 10^{3\div 4}GeV$ the 5D scale $M$ is of
order $(\mu M^2_{pl})^{1/3}\sim 10^{14}GeV$.

At times $t\sim t_0$ such that $H_0 (t_0 -t)\ll 1$
solution of (\ref{r7}) is the same as in the standard cosmology
\be
\l{a11}
z\simeq 1+H_0 (t_0 -t).
\ee 
In the region (ii) where $y\ll 1$, from (\ref{r2}) we obtain
\be                                                         
\l{a12}                                                     
u(t) \simeq                                                 
z^{-2} \simeq (H_0 t)\left(\frac{6 +2(1+q_0 )r^2 }{r^2}\right)^{1/2}.
\ee

In regions (ii) and (iii) solution of Eq. (\ref{r4}) is
\be
\l{r8}
u^2 (t)=\frac{2(3+(1+q_0 )r^2 )}{r^2} (H_0 t)^2 +\frac{4 \r_{r0}}{
r^2\,\mu \r_{m0}}H^2_0 t.
\ee
The transition time $\bar{t}$
 from the law $z\sim (H_0 t)^{-1/4}$ to $z\sim H_0 (t)^{-1/2}$ is
\newline $\bar{t} \sim 10^{-(7\div 8)}(3+(1+q_0 )r^2 )^{-1}GeV^{-1}$. 
This time is considerably
smaller than the characteristic times of nucleosynthesis  
 $1\div 10^2 s$, or
$10^{24\div 26}GeV^{-1}$. Thus, at the time of nucleosynthesis we can
neglect
second term in (\ref{r8}). 

Let us compare primordial nucleosynthesis in the standard and non-standard
cosmologies.

First, let us find relation between time and temperature of the
Universe in the radiation-dominated phase in the non-standard cosmology.
At $z>z_{cr}$ the Universe becomes radiation-dominated.
Combining expressions for the radiation density
$$
\h{\r}_r (z)\simeq \h{\r}_{r0}z^4
$$
 valid for large $z$  and the expression for the radiation density as a
function of temperature $\t{T}$ of the Universe \cite{kolb,wein}
$$
\h{\r}_r (\t{T}) = \frac{\pi^2}{30}{g}_{*}(\t{T}) \t{T}^4
$$
and substituting $z^4$ from (\ref{r8}) we have
\be
\l{a14}
\frac{\pi^2}{30}{g}_{*}(\t{T})  \t{T}^4 
\simeq\frac{\h{\r}_{m0}}{4(H_0 t)^2}\,\,
\frac{2r^2}{3+(1+q_0 )r^2 }\frac{\h{\r}_{r0}}{\h{\r}_{m0}}
.\ee
In the standard cosmology, the Friedman equation
in the radiation-dominated period
$$
\left(\frac{\dot{a}(t)}{a(t_0 )}\right)^2 =\frac{\k^2_4
\h{\r}_{r0}}{3}\left(\frac{{a}(t)}{a(t_0 )}\right)^2   
$$
is solved as
$$ z^{-2} =2t\left(\frac{\k^2_4 \h{\r}_{r0}}{3}\right)^{1/2}
,$$
where we substituted ${a}(t_0 )/a(t )\simeq z$.
Substituting the relation
$$
\frac{\k^2_4 \h{\r}_{r0}}{3}=\frac{\k^2_4 \h{\r}_{m0}}{3}
\frac{\h{\r}_{r0}}{\h{\r}_{m0}}=
H^2_0 \frac{\h{\r}_{r0}}{\h{\r}_{m0}}\Omega_m
$$
in the expression for radiation density $\h{\r}_r (z)\simeq \h{\r}_{r0}z^4$,
we obtain
$$
\h{\r}_r (z)\simeq\frac{\h{\r}_{m0}/\Omega_m}{4(H_0 t)^2}.
$$
The relation connecting time and temperature takes the form
\be
\l{a141}
\frac{\pi^2}{30}g_{*}(T) T^4 \simeq\frac{\h{\r}_{m0}/\Omega_m}{4(H_0 t)^2}.
\ee
Comparing  expressions (\ref{a14}) and (\ref{a141})
we notice the appearance of the large extra factor  
$\h{\r}_{m0}/ \h{\r}_{r0}\simeq 0.5\cdot 10^{4}$ \cite{part} in the
non-standard cosmology . 
The effective numbers of massless degrees of freedom ${g}_{*}(\t{T}) $
 and $g_{*}(T )$
are temperature-dependent and are different in the standard and
non-standard cosmologies.

Time dependence of the Hubble parameter in both the
standard and
non-standard cosmologies is the same $H(t)=1/2t$.
The Hubble parameter expressed as a function of temperature in the
non-standard cosmology is
$$
H({\t{T}})={\t{T}}^2 \left( \frac{\pi^2 {g}_{*}(\t{T}) }{30}
\frac{H_0^2}{\h{\r}_{m0}}
\frac{[3+(1+q_0 )r^2 ]}{2r^2}\frac{\h{\r}_{m0}}{\h{\r}_{r0}}\right)^{1/2}.
$$ 
The freezing temperature  $\t{T}_F$
of the reaction $n\leftrightarrow p$ is estimated as the temperature
 at which the
Hubble parameter $H(\t{T})$ is of order of
the reaction rate  $\sim G^2_F\t{T}^5$ \cite{wein,kolb}.
Comparing the freezing temperatures $T_F$ and $\t{T}_F$ in
the standard and non-standard cosmologies we obtain
\be
\l{t}
\frac{\t{T}_F}{T_F}=\left(\frac{{g}_{*}(\t{T}_F) }{{g}_{*}(T_F )}
\frac{[3+(1+q_0 )r^2]}{2r^2\Omega_m }
\frac{\h{\r}_{m0}}{\h{\r}_{r0}}\right)^{1/6}
.\ee
At the freezing
temperature $\t{T}_F$ the effective number of degrees of freedom is
${g}_{*}(\t{T}_F )=10.75$. At temperatures below 1 $MeV$ neutrinos
decouple from massless particles in equilibrium and the effective   
 number of degrees of
freedom is ${g}_{*} (T_F)\sim 5$ yielding
the factor $\left({{g}_{*} (\t{T}_F )}/ g_{*} (T_F )          
\right)^{1/6} \sim 1.1$.  
Taking for an estimate $r^2 =1;\,\,10;\,\,2\cdot 10^4$  and $q_0 =0$, 
 we obtain
$$
\frac{\t{T}_F}{T_F}\simeq 7.9;\,\,6.5;\,\,5.6 . 
$$
Substituting $ T_F\simeq 1 MeV$,  we obtain the equilibrium ratio  
$$
(n/p)(\t{T}_F )= \exp{\left(-\frac{m_n -m_p }{\t{T}_F}\right)}
\simeq 0.80; \,\,0.76;\,\,0.72
$$
as compared with $(n/p)({T}_F )\simeq 0.17$ in the standard cosmology.
 
The mass fraction of ${}^4$He produced in the  nucleosynthesis $X_4
=4n_{He}/n_N$  starts rapidly approaching the final value at 
the temperature $T_f\sim 0.1 - 0.3 MeV$ at the time $t_f$.   
The times of the freeze-out
in the standard and non-standard cosmologies are $t_F \simeq 0.75s$ and
$\t{t}_F \sim 10^{-4} s$.
In the standard cosmology the time between the freezing point and $t_f$
is about 2 minutes during which the ratio $n/p$ decreases because of the
neutron decay \cite{wein,kolb}. In non-standard cosmology the time 
$\t{t}_f$ is of order a
second and for the estimate of $X_4$ the neutron decay between the times
$\t{t}_F$ and $\t{t}_f$ can be neglected.   

As a result, in the non-standard cosmology the mass fraction of  ${}^4$He
calculated with the above values of $(n/p)$ is 
$$ X_4 = \frac{2(n/p)}{(n/p) +1}\simeq 0.89;\,\,0.86;\,\,0.84
$$
which is to be compared with $X_4 \simeq 0.25$ in the standard cosmology
\cite{kolb,part}.

Assuming that there is a large hierarchy of scales, above we considered the
case $b^2 \geq 1$. However, because of the large value of the ratio
$M_{pl}/\m$, there can be a hierarchy between $M$ and $M_{pl}$ with $b^2 <
1$, supposing that $b^2$ is not too small. 

Let us estimate the age of the Universe and the freezing temperature of 
the reaction $n\leftrightarrow p$ assuming that $b^2 < 1$ or $r^2 >1$.
 We will
distinguish two cases (i) $q_0 \neq -1$, and (ii) $q_0\simeq -1$.
In the first case, taking $|q_0|= O(1)$ \cite{part, wein}, with large $r^2
\gg 1$ 
we obtain that  $2r^2\Omega_m /(3+(1+q_0 )r^2 ) =O(1)$. As it was discussed
above, in (\ref{t})
there is a big number $\h{\r}_{m0}/\h{\r}_{r0}\sim 2\cdot 10^4$ which for
large $r^2$ yielded $X_4 \sim 0.8$.  
We see that
 $X_4$ weakly depends on $r^2$, and its value is about three times larger
than that in the standard model. 

In the case (ii) the ratio of freezing temperatures (\ref{t}) contains the
factor $2r^2\Omega_m /3$  which for $r^2 \sim 10^4$ is of
order unity, i.e. $\t{T}_F\sim {T}_F$.  The estimate of the age of
the Universe is
$$
H_0 t_0 =\frac{r}{\sqrt{2}}\int^1_0 \frac{du}{\left[(2r^2 +1)u^2 +3 -4u^{1/2}
\right]^{1/2}}\simeq \frac{1}{2}\int^1_{\sqrt{3/2r^2}} \frac{du}{u}
.$$ 
With $r^2\sim 10^4$ the age of the Universe is obtained considerably above the
experimental bound.    

In the model with 5D action  containing explicit curvature terms on the
branes (on the visible brane with the Newton constant) situation is rather
different. In this model in the range of times at which $H^2 (t)/\m^2 \ll 1$
 solution of the
generalized Friedman equation is close to solution of the Friedman equation
in the standard cosmology with matter (radiation) and cosmological constant
of the form $H^2/H^2_0 =\Omega (z+1)^n + (1-\Omega )$, 
where $\Omega =\Omega_m\, (\Omega _r ) , n=3 \,(4) $ in the cold matter
(radiation) dominated Universe (cf. \cite{Shtanov2,Shtanov3,Def}).
 With the input of the current values of the
Hubble parameter and matter density, in terms of the cosmological parameter
$z$, the above range is estimated as $z< 10^{22} $.

\end{document}